\documentclass[conference]{IEEEtran}
\IEEEoverridecommandlockouts

\usepackage{etoolbox}

\newbool{showcomments}
\setbool{showcomments}{false} 

\newcommand{\ewa}[1]{\ifbool{showcomments}{\textcolor{blue}{[E: #1]}}{}}
\newcommand{\vch}[1]{\ifbool{showcomments}{\textcolor{green}{[V: #1]}}{}}
\newcommand{\gautier}[1]{\ifbool{showcomments}{\textcolor{red}{[G: #1]}}{}}

\newcommand{\figref}[1]{\hyperref[#1]{Fig.~\ref{#1}}}
\newcommand{\secref}[1]{\hyperref[#1]{Sect.~\ref{#1}}}
\newcommand{\tabref}[1]{\hyperref[#1]{Tab.~\ref{#1}}}

\usepackage{cite}
\usepackage{amsmath,amssymb,amsfonts}
\usepackage{graphicx}
\usepackage{textcomp}
\usepackage{xcolor}
\def\BibTeX{{\rm B\kern-.05em{\sc i\kern-.025em b}\kern-.08em
    T\kern-.1667em\lower.7ex\hbox{E}\kern-.125emX}}
\usepackage{tikz}
\usepackage{tikz-3dplot-circleofsphere}
\usetikzlibrary{angles,quotes,patterns}
\usepackage{arydshln}
\usepackage{svg}
\usepackage{tcolorbox}
\usepackage{tikz}
\tcbuselibrary{skins}
\usetikzlibrary{shapes.arrows, positioning}
\usepackage{algpseudocode}
\usepackage{array}
\usepackage{multirow} 
\usepackage{colortbl}
\usepackage{rotating}
\usepackage{lineno} 
\usepackage[T1]{fontenc} 
\usepackage[cmintegrals]{newtxmath}
\usepackage{bm} 
\usepackage{booktabs}
\usepackage[colorlinks=true,linkcolor=red,citecolor=blue]{hyperref}
\usepackage[hypcap=true]{caption}
\usepackage[font=small]{caption}
\usepackage{wrapfig}

\DeclareMathOperator*{\argmax}{argmax} 

\usepackage{mathtools, stmaryrd}
\usepackage{xparse}
\usepackage{cite}
\DeclarePairedDelimiterX{\Iintv}[1]{\llbracket}{\rrbracket}{\iintvargs{#1}}
\NewDocumentCommand{\iintvargs}{>{\SplitArgument{1}{,}}m}
{\iintvargsaux#1} %
\NewDocumentCommand{\iintvargsaux}{mm} {#1\mkern1.5mu..\mkern1.5mu#2}

\begin{document}

\title{SWIFT: Semantic Watermarking for Image Forgery Thwarting}
\author{
    Gautier Evennou$^{1,2}$,
    Vivien Chappelier$^{2}$,
    Ewa Kijak$^{1}$,
    Teddy Furon$^{1}$
    \\  \\
    $^1$\textit{IRISA, Univ. Rennes, Inria, CNRS } \qquad 
    $^2$\textit{Imatag} 
}

\maketitle

\begin{abstract}
    This paper proposes a novel approach towards image authentication and tampering detection by using watermarking as a communication channel for semantic information. We modify the HiDDeN deep-learning watermarking architecture to embed and extract high-dimensional real vectors representing image captions. Our method improves significantly robustness on both malign and benign edits. We also introduce a local confidence metric correlated with Message Recovery Rate, enhancing the method's practical applicability. This approach bridges the gap between traditional watermarking and passive forensic methods, offering a robust solution for image integrity verification.
\end{abstract}


\def\Hide{Hide-$\mathbb{R}$ }

\section{Introduction}
\label{sec:intro}
Many technical means can verify the authenticity of multimedia content.
This ranges from a digital signature stored in the metadata like in the recent C2PA (Coalition for Content Provenance and Authenticity) and IPTC (International Press Telecommunications Council) initiatives, to passive forensics~\cite{chen2008determining,farid2016photo} and active fragile watermarking~\cite{Celik,Fridrich}.
The main difficulty resides in making a clear cut between benign processing which are common editing in the entertainment industry and malicious transformations which modify on purpose the content.
Semi-fragile watermarking faces this challenge: it should be robust to benign processing but fragile to deeper transformations.
At the decoding side, its absence reveals that the piece of content has been modified beyond the accepted limit.
This limit between benign and malicious editing is not easy to be defined in mathematical terms, although in real life the difference is straightforward:
Any modification of the semantics is malicious.

This paper investigates the idea of hiding semantics information within the cover work in an imperceptible and robust manner.
The verification amounts to compare the semantics of the content with the decoded information.
To the best of our knowledge, embedding its own semantic into the content itself to ensure integrity is an unexplored research path. 
A first challenge lies in the poor capacity of robust image watermarking.
Multi-bit watermarking embeds messages into images but typically only supports up to 64 bits of data transmission.
Higher capacity schemes exist but with a much lower robustness.
The second challenge is the representation of the semantic of an image whose definition is still a matter of debate.
We chose the textual description of the image given by an automatic captioning as the message to be hidden.

The scenario establishes a covert channel between two entities: Alice, the sender who authenticates the original work, and Bob, the recipient tasked with verifying its authenticity.
The cover work may undergo modifications by a third party, referred to as Eve, acting as an intermediary.
Eve's alterations may be intentional, involving semantic edits, or unintentional, comprising benign changes.
The crux of our method lies in Bob's ability to recover the message embedded by Alice.
This recovery enables Bob to assess whether Eve's modifications have introduced semantically misleading alterations to the original content.
By comparing the recovered message with the received work, Bob makes informed decisions about the nature and extent of any change.
The robustness of the communication channel despite potential interferences is key.

To this end, we propose to increase the utility, re-usability and flexibility by disentangling the watermarking layer from the encoding layer.
The watermarking layer is responsible for hiding a high-dimensional real-valued unit-norm vector in the cover while optimizing robustness to various transforms and the watermark imperceptibility. The encoding layer is responsible for encoding a message as a signal to be transmitted on this noisy communication channel.
The decoding layer then retrieves the message with some confidence level.

This framework provides a robust mechanism for authentication and content verification in scenarios where the integrity of digital media may be compromised between creation and reception.
This paper introduces three contributions: 
\begin{itemize}
    \item \Hide : Inspired by HiDDeN~\cite{Zhu2018HiDDeNHD}, we propose an encoder-decoder network architecture jointly trained to embed and extract high-dimensional unit-norm vectors in images.
    \item Encoding layer: It encodes a variable-length binary message into a vector to be hidden in images.
    \item Caption Compression: We finetune a large language model for captioning and combine it with an arithmetic codec to compress the payload as in LLMZip~\cite{LLMZip}.
\end{itemize}
Three major features stem from the combination of these contributions into the SWIFT scheme:
\begin{itemize}
    \item Reliability: A confidence metric on the decoded caption gives an informed decision-making about authenticity.
    \item Security: The design guarantees security via a secret key.
    \item Performance: SWIFT achieves state-of-the-art results across various benign and malicious transforms, demonstrating its robustness in challenging scenarios.
\end{itemize}

\section{Related Work}

\paragraph{Image forensics}
Passive methods detect alterations of a piece of content, possibly malicious ones.
They utilize noise residuals or high-frequency features as input to highlight manipulation traces.
These methods are limited to providing localized insights into \emph{specific} alterations.
For instance, copy-move forgery detection uses Siamese networks~\cite{wu2018busternet} while splicing detection leverages two-stream architectures~\cite{zhou2018learning, wu2019mantra}.
Inpainting detection methods have focused on traces left by some deep inpainting models~\cite{zhu2018deep,li2019localization}.
Forensics methods lack the capacity to offer a global perspective due to the absence of contextual information from the original image.

\paragraph{Image watermarking}
Traditional watermarking schemes embeds invisible marks within multimedia content to assert copyright ownership (robust watermarking) or authenticate content (fragile watermarking).
Classic techniques involve manipulating spatial or frequency domains representation of the media~\cite{IH99,cox2007digital}.
Recently, deep-learning enabled more robustness as first shown with the encoder-decoder HiDDeN architecture~\cite{Zhu2018HiDDeNHD} and followed with~\cite{vukotic2018deep,zhang2019robust}.
SSL~\cite{ssl} embeds a binary message in the latent space of a foundation model learned with supervised learning with low perceptibility but high inference cost due to its iterative nature.
TrustMark~\cite{bui2023trustmarkuniversalwatermarkingarbitrary} leverages a more classic encoder-decoder architecture and a GAN loss to learn how to embed binary messages.
Note that the payload of a watermarking scheme is always fixed in the literature. One of our contributions is to tackle variable-length messages.

Watermarking can be used for authenticity verification, but it usually uses a fragile or semi-fragile signal whose absence reveals tampering~\cite{Celik,Fridrich}.
One exception is the idea of embedding a compressed representation of the image in itself with robust watermarking~\cite{10.1007/978-3-642-40099-5_11}.
At the detection stage, the verifier finds back a copy of the original image to be compared with the image.
Our work is similar in spirit except that we embed the semantic textual description of the original image.

\section{Method}
\begin{figure*}[t!]
    \centering
    \includegraphics[width=0.9\linewidth]{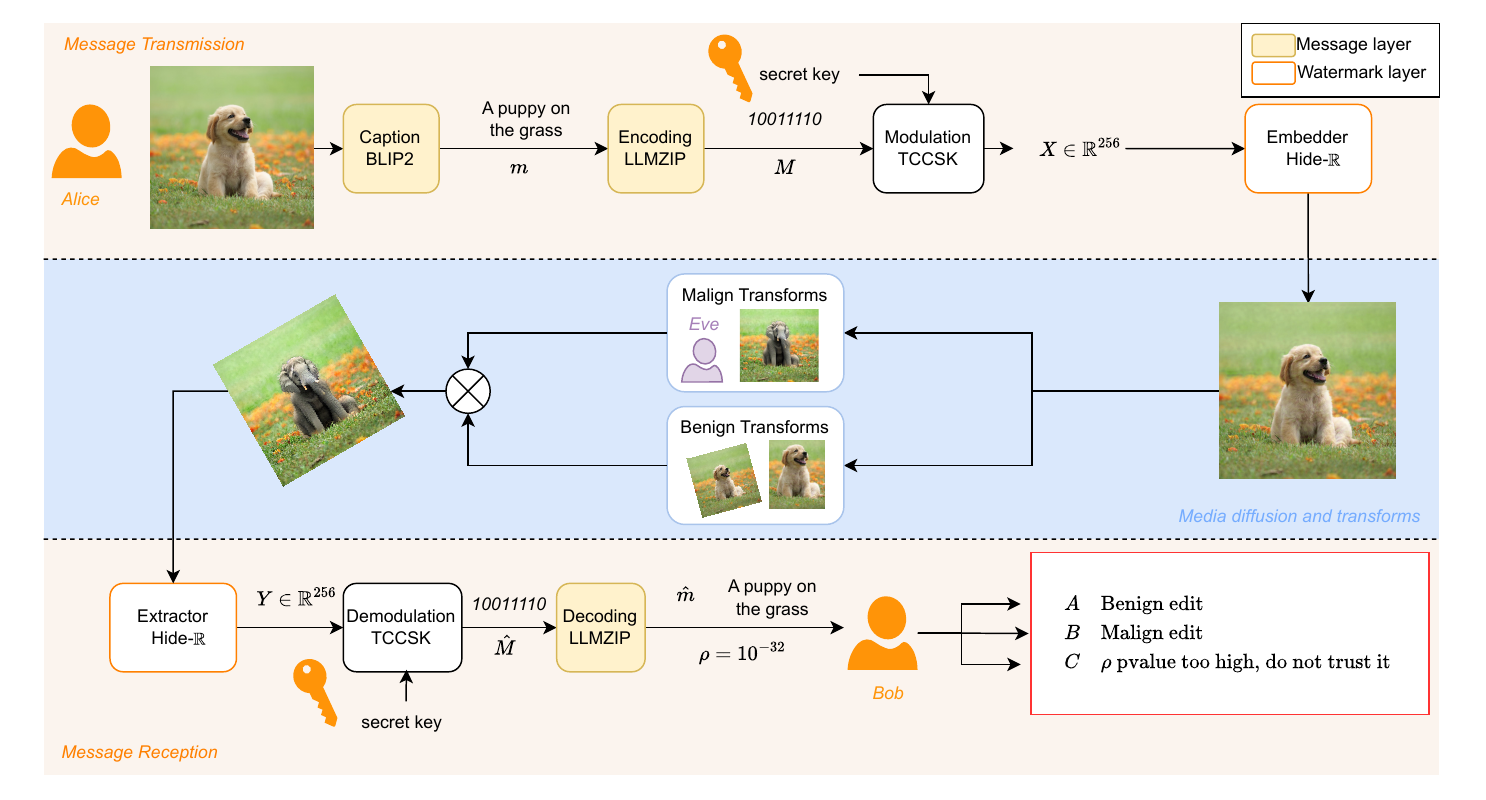}

    \caption{Overview of SWIFT. To ensure the integrity of an image, Alice first leverages the message layer to tailor a semantic representation of the image, in our case a caption compressed in a lossless fashion. The resulting bit stream is fed to the TCCSK modulation layer thus enabling security and confidence (see \secref{sec:decoding_confidence}), and then to the watermarking layer based on our \Hide encoder-decoder neural network. At reception, Bob executes the inverse process with a secret key and obtains both the caption and the p-value $\rho$. If $\rho$ is low enough, Bob can entrust the decoded caption and compare the image he received with the caption, enabling comparison between a proxy of the original image and the received one. 
    }
        \label{fig:pipeline}

\end{figure*}

This section presents the design of the encoding and watermarking layers. We break down the encoding layer into two primary components: the message layer and the modulation layer. ~\hyperref[fig:pipeline]{Fig.~\ref{fig:pipeline}} depicts our method.

\subsection{The message layer}

Alice wants to transmit a message $M$ to Bob so that he can assess the integrity of the cover image.
Alice uses an image captioning model like BLIP2~\cite{li2023blip2} to generate the caption $m$ of the cover. 

Alice uses arithmetic coding~\cite{arithmeticENCDEC} for losslessly compressing $m$ into $M$ to reduce the number of bits.

As in LLMZip~\cite{LLMZip}, Alice takes advantage of a LLM to model the distribution of the messages and improve the compression.
She uses OPT-125m~\cite{Zhang2022OPTOP} finetuned on BLIP2 captions from $2,000$ MSCOCO validation set images.
This acts as oracle and gives the probability of each caption symbol used by the arithmetic coding~\cite{Huang2024CompressionRI}.

\begin{figure}
    \centering
    \includegraphics[width=\linewidth]{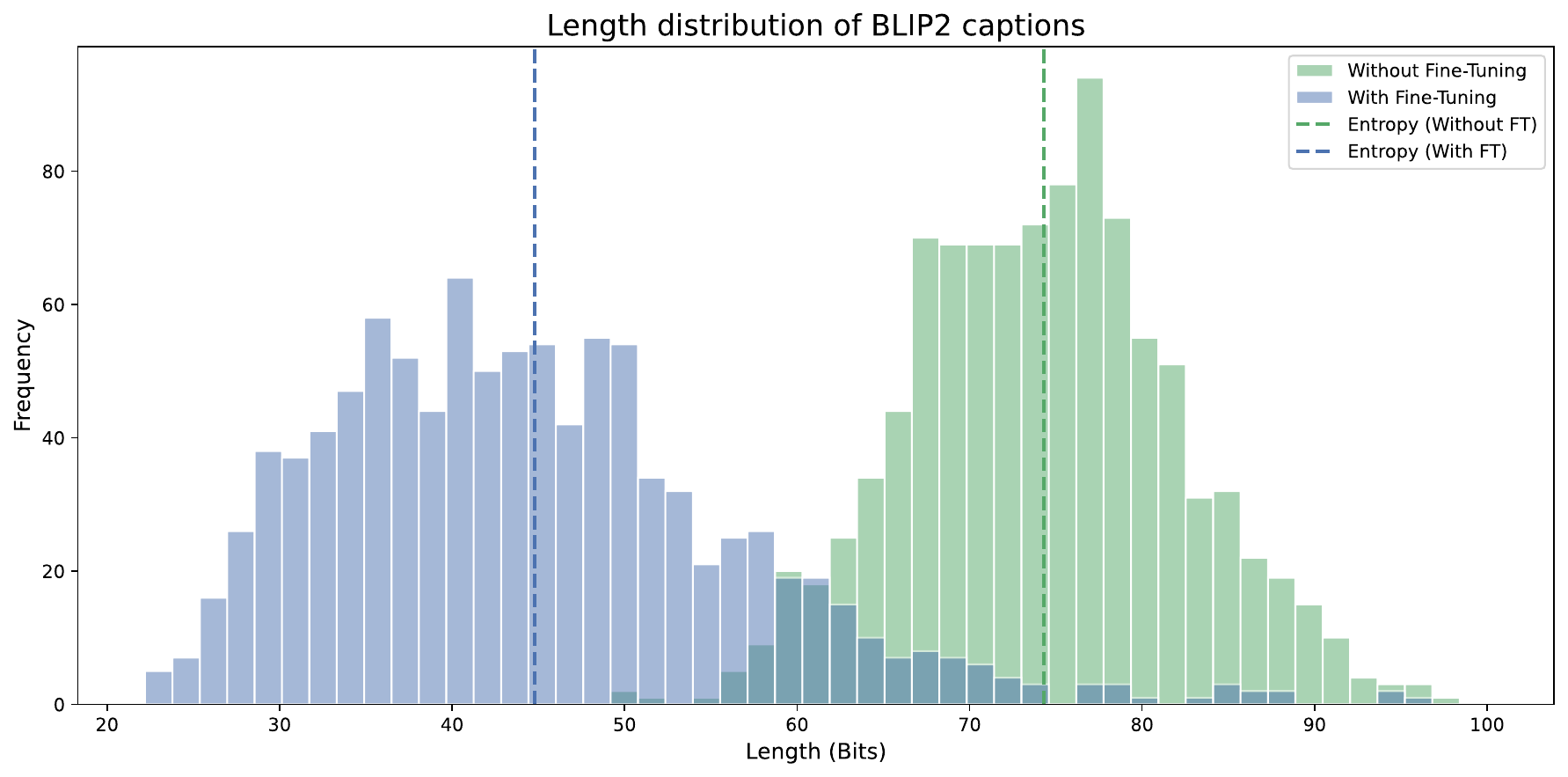}
    \caption{Length distribution of BLIP2 captions encoded by OPT-125m version. We show that the finetuned version leads to entropy reduction and thus is more efficient to encode captions.}
    \label{fig:entropy_distrib}
\end{figure}

\figref{fig:entropy_distrib} shows that finetuning OPT on BLIP2 reduce the mean capacity needed from $75$ to $45$ bits. 

\subsection{The watermarking layer}
\begin{figure}[h!]
    \centering
    \includegraphics[width=0.9\linewidth]{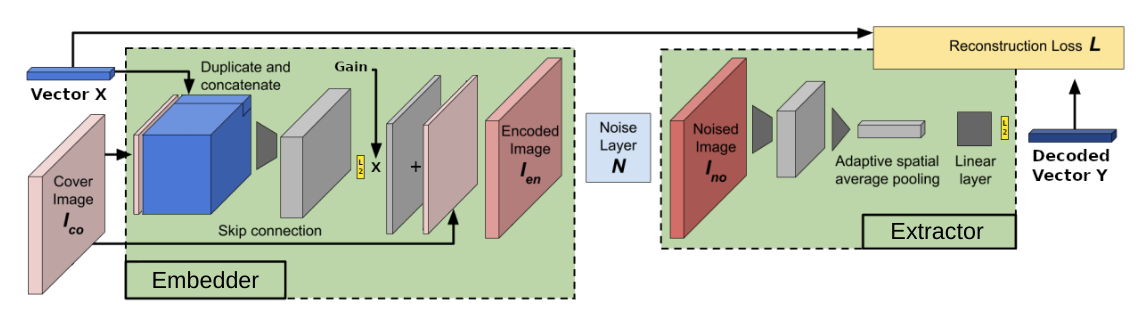}
    \caption{
    \Hide architecture. We use a L2 norm to control the watermark power to enforce a target PSNR on both the watermark signal in the embedder and on the decoded vector in the extractor.}
    \label{fig:hidden}
\end{figure}

Modern watermarking leverages deep-learning to learn end to end how to embed a message into a cover image.
It enables robustness against benign edits by performing augmentations between the watermark embedding and watermark extraction stages~\cite{Yang2022ImageDA}.
The most famous example is HiDDeN~\cite{Zhu2018HiDDeNHD} based on two convolutional neural networks (CNN) jointly trained to embed and extract a fixed-length binary message.

\figref{fig:hidden} depicts our \Hide architecture, resulting from several modifications of HiDDeN.
The input data works with unit vectors in high-dimensional real space instead of binary messages.
Specifically, we draw random samples $X$ uniformly distributed on the surface of the unit hypersphere in $\mathbb{R}^D$ with $D = 256 $ :
\begin{equation}
\label{eq:rng}
X = \frac{Z}{\|Z\|} \quad\text{with}\quad Z \sim \mathcal{N}(0,I_D).
\end{equation}
We extend the number of channels in the convolutional layers to $1.5D$ instead of the fixed $64$ to account for higher dimension $D$ than the message length $L=30$ proposed in the original paper. Signal $X$ is concatenated with the cover image $I_{co}$ along the channel dimension before the first convolutional layer.
Instead of using a discriminator, we opt for a fixed PSNR budget which both enforce imperceptibility in a flexible way and speed up the learning process.
The training minimizes the reconstruction loss $\|X-Y\|$ between $X$ and the reconstructed unit vector $Y$.

This framework gives a zero-bit watermarking system. Assume $X_0$ is drawn according to (\ref{eq:rng}) from a pseudo-random generator seeded by the secret key $K$ associated with a fixed index $M=0$. A watermarking signal is deemed present if the cosine similarity $C_0=Y^\top X_0$ is above a threshold $c$. Under the hypothesis $\mathcal{H}_0$, the cover is not watermarked or watermarked with another secret key $K'$.
Then, the p-value is defined as the probability of having higher cosine similarity $C_0$ than the threshold $c$ and given by:
\begin{equation}
\label{eq:rho0}
\rho_0(c) = P(C_0 \geq c) = 
\begin{cases}
      \frac{1}{2} I_{1-c^2}\left(\frac{D-1}{2},\frac{1}{2}\right), & \text{if}\ c > 0 \\
      \frac{1}{2} I_{c^2}\left(\frac{1}{2},\frac{D-1}{2}\right), & \text{otherwise,}
    \end{cases}
\end{equation}
where $I$ is the regularized incomplete beta function.
This illustrates how a confidence value is available to Bob at the watermark extractor.

\subsection{The modulation layer}  
\subsubsection{Multi-bit watermarking with confidence}

One way to use zero-bit watermarking to send a $N$-bit message $M$ is to share $2^N$ different secret keys, each associated with one possible message.
Then Alice selects the key $K$ corresponding to the message to send, and Bob runs the watermark extractor with all the $2^N$ possible keys. Bob selects the decoded message as:
\begin{equation}
\hat{M} = \argmax_{m \in \Iintv{0,2^N-1}}(C_m)
\end{equation}

The p-value under $\mathcal{H}_0$ of decoding this specific message $\hat{M}$ by chance in a non-watermarked image is given by:
\begin{equation}
\rho_1(c) = 1-(1-\rho_0(c))^{2^N} \lesssim 2^N \rho_0(c) \text{ if } 2^N \rho_0(c) \ll 1
\label{eq:tradeof}
\end{equation}

Equation~\eqref{eq:tradeof} reveals a trade-off between minimizing $\rho_1(c)$ and increasing the length $N$ of the message:
If $N > -\log_2(\rho_0(c))$, it is likely that the decoded message cannot be trusted by Bob.
This gives an approximation of the maximum quantity of information that may be transmitted.

\subsubsection{Modulating variable-rate signals}
\begin{figure}
    \centering
    \includegraphics[width=\linewidth]{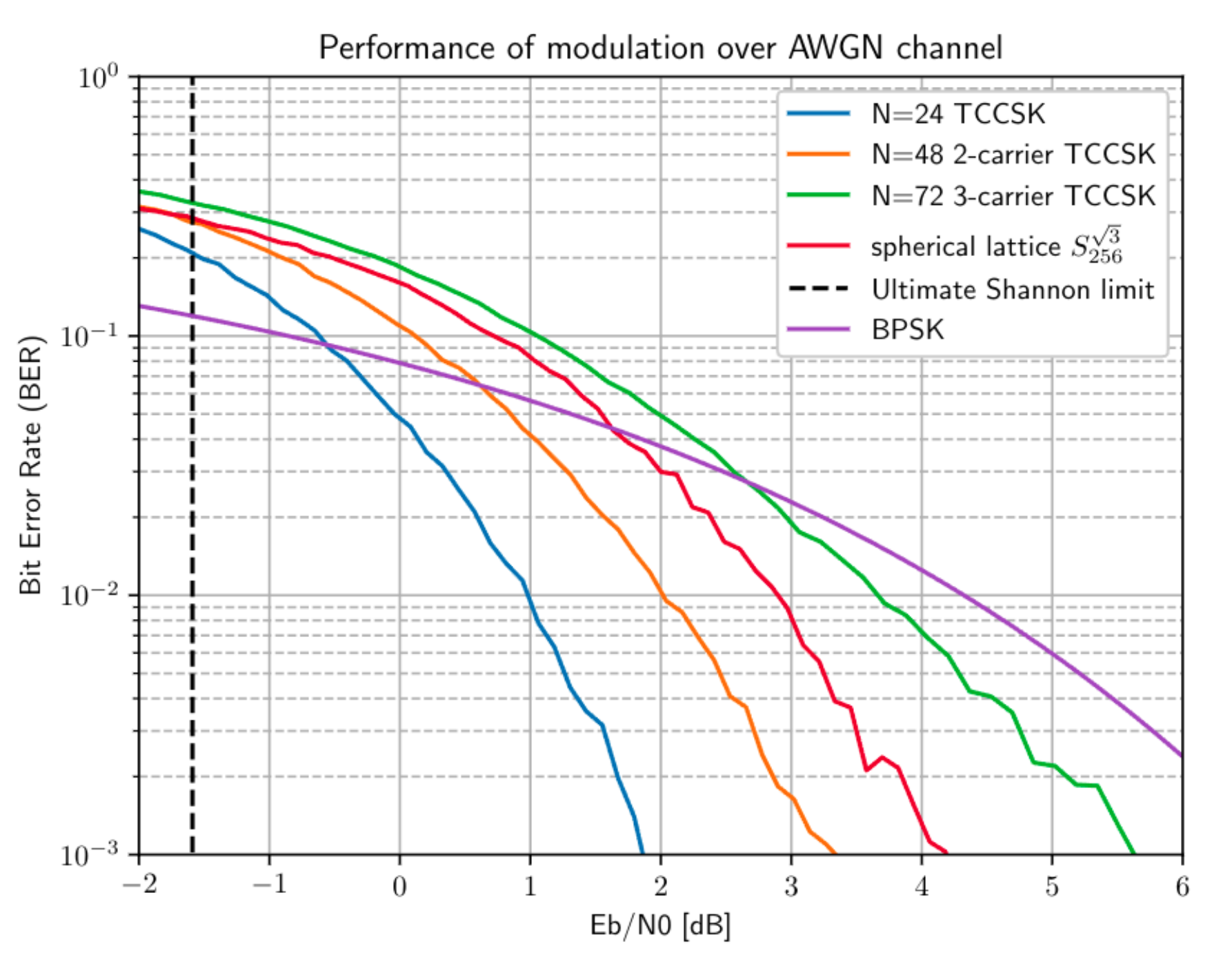}
    \caption{Performance of different encoding schemes under additive white Gaussian noise. $S_{256}^{\sqrt{3}}$ is the spherical lattice from \cite{spreadingvectors}, capable of encoding $22108160 \simeq 2^{24.4}$ messages. TCCSK is most suited for our use case as it can cope with variable length messages. 
}
    \label{fig:ber}
\end{figure}

In the previous section, Bob needs to test  $2^N$ keys which becomes intractable as $N$ grows.
An alternative is to generate multiple carriers from the same pseudo-random generator seeded by secret key $K$, and to combine them to produce $X \in \mathbb{R}^D$. 
This paper uses truncated cyclic shift keying (TCCSK) modulation~\cite{1238736} for its improved performance in the additive white Gaussian noise (AWGN) channel compared to Binary Phase Shift Keying (BPSK) (see Fig.~\ref{fig:ber}).
The message $M$ is split into $T$ equal blocks of length $L$, padding the last block with $0$ if necessary.
Then the $j$-th block is encoded from a carrier $Z_j \in \mathbb{R}^{2^L} \sim \mathcal{N}(0,I_{2^L}), j \in \Iintv{0,T-1}$ by cyclically shifting it by a number equal to the value represented by the $L$-bit sub-message $M_j \coloneqq M_{\Iintv{jL,(j+1)L}}$. The carriers are truncated to dimension $D$, normalized, summed, and normalized again for transmission:
\begin{equation}
X = \frac{Z}{\|Z\|} \quad\text{with}\quad  Z = \sum_{j=0}^{T - 1} \frac{Z_j^{M_j}}{\|Z_j^{M_j}\|},
\end{equation}
where $Z^a$ denotes a cyclic shift of $Z$ by $a$ followed by truncation to the first $D$ dimensions.
Decoding is achieved by recovering each sub-message from:
\begin{equation} 
\label{eqn:demodulation}
\hat{M}_j = \argmax_{k \in \Iintv{0,2^L-1}}(Y^\top\frac{Z_j^k}{ \|Z_j^k\|}).
\end{equation}
The p-value $\rho$ is then obtained from Fisher's combined probability test \cite{Fisher1992} on evaluations of (\ref{eq:tradeof}) for each submessage: 
\begin{equation} 
\label{eqn:fisher}
\rho = 1-\gamma(T,-\sum_{j=0}^{T - 1} \log(\rho_1(C_j)))\text{, with } C_j = Y^\top \frac{Z_j^{\hat{M}_j}}{\|Z_j^{\hat{M}_j}\|},
\end{equation}
where $\gamma$ is the lower incomplete gamma function.
Although we would ideally want to set $L$ to $N$, testing all $2^L$ cyclic shifts becomes untractable as $L$ grows.
Proceeding by blocks tackles the variable-length of the message.
Yet, some capacity is lost due to padding unless the message length $N$ is a multiple of $L$. 
Finally, to illustrate the flexibility of the approach, we also compare TCCSK with lattice-based modulation on the spherical lattice used in \cite{spreadingvectors}. Although this method is very fast and better than BPSK for the $24$-bit case, its efficiency is lower than TCCSK with a single carrier (see \figref{fig:ber}). 

\subsection{Security}
Our approach adheres to Kerckhoffs's principle, relying solely on the shared secret key $K$ between Alice and Bob for security. We assume Eve has full knowledge of the system, except for $K$. In content authentication, the attacker's primary goal is to forge watermarked content without $K$ (spoofing attack), rather than removing existing watermarks. Each system use reveals at most $D$ out of $2^L$ carrier values, reducing the urgency for key rotation if usage is limited.
As a symmetric system, both Alice and Bob can produce forgeries using $K$. Thus, mutual trust between them is assumed, with Eve being the only untrusted party in our threat model.

\section{Experiments \& Results}

This section introduces evaluation of SWIFT and recent watermarking methods for the task of message recovery. 
\subsection{Evaluation}

\noindent\textbf{Metrics.} The benchmark compares watermarking methods by their Message Recovery Rate (MRR), which is defined as the rate of messages being perfectly transmitted without any modifications, over a test set of watermarked images $\mathcal{I}_t$. 
Let $m_i$ be the original caption and $\hat{m_i}$ the corresponding recovered caption for an image $I_i \in \mathcal{I}_t$. The MRR is defined as follows:
\begin{equation}
MRR = \frac{1}{|\mathcal{I}_t|}\sum_{i=1}^{|\mathcal{I}_t|} \delta( m_i, \hat{m}_i)
\end{equation}

We chose this metric to ensure practicability and accurately assess the robustness of a system. A watermarking system designed for our task should strive to reach $100\%$ MRR, especially when no confidence metric is available at the decoding step, which is the case of all systems but SWIFT. 

\noindent\textbf{Test set.} Our test set $\mathcal{I}_t$ is composed of 20,220 images. We use the Emu Edit test set \cite{Sheynin2023EmuEP} which comprises 2,022 images from MSCOCO \cite{chen2015microsoftcococaptionsdata} and editing instructions for Image Editing models from 8 classes (\textit{local, add, remove, global, text, background, style, color}).
For each image, we perform 6 benign and 1 malign transformation with four variations in classifier-free guidance. Benign ones are chosen to be realistic in a web setting, or quite important distortion-wise but without semantic alterations: crop 40\% of image surface, random noise, grayscale conversion, resize to $128 \times 128$, jpeg compression with quality coefficient at 50. Malign ones are images edited by a diffusion model according to Emu Edit instructions, supposed to change the meaning of the cover work.

\subsection{Comparison with state of the art}
Table~\ref{tab:zero_ber_performance} shows the results of state-of-the-art methods SSL and TrustMark against SWIFT: we observe superior resilience to malign transforms while maintaining state-of-the-art performance on benign modifications with significant improvement on resize and grayscale transforms due to our training. We provide another version of SWIFT to watermark at 42db which performs better than TrustMark(Base) on almost all settings.

\begin{table*}
\centering
\caption{Message Recovery Rate by Transform Type. All the methods share the same messages from the LLMZip as inputs. We provide results for SWIFT at 2 different target PSNR for fair comparison with existing methods.}
\label{tab:zero_ber_performance}
\begin{tabular}{>{\centering\arraybackslash}p{2.3cm} >{\centering\arraybackslash}p{2.3cm} >{\centering\arraybackslash}p{2.3cm}>{\centering\arraybackslash}p{2.3cm} >{\centering\arraybackslash}p{2.3cm} >{\centering\arraybackslash}p{2.3cm}>{\centering\arraybackslash}p{2.3cm}}
\toprule
\multirow{2}{*}{Transform} & \multicolumn{5}{c}{Message Recovery Rate(\%) \(\uparrow\) } \\
\cmidrule(lr){2-6}
 & SWIFT& SWIFT & SSL & TrustMark(Q) & TrustMark(Base) \\
 PSNR(db) & 40 & 42 & 40 & 42.5 & 41.6 \\
\midrule
global & \textbf{63.7}& 45.1 & 0.45 & 0.0 &39.0\\
text & \textbf{65.3}& 45.9 &0.38 &0.0&40.0\\
style & \textbf{64.8} & 45.3 &1.32 &0.0&38.3\\
local & \textbf{65.9} & 48.2 &0.39 &0.0&40.2\\
background & \textbf{66.5} & 48.3 & 0.75 &0.0&43.9\\
color & \textbf{65.3}& 46.9 &1.53 &0.0&44.6\\
remove & \textbf{67.2}& 50.0 & 0.0 & 0.0 &48.0\\
add & \textbf{69.5}& 49.2 &0.0 & 0.0&43.7\\
\midrule
crop 40\% & 82.4&76.7&  50.1 & 86.4 & \textbf{91.9}
 \\
noisy & \textbf{75.5}&65.4 & 0.0 & 11.8 & 17.2 \\
resize 128 & \textbf{95.1}&91.3& 0.00 & 0.0 & 0.0 \\
grayscale & \textbf{95.8}& 93.8& 2.82 & 78.0&91.2\\
jpeg@50 & \textbf{96.2} & 93.5&1.5 & 82.1 &93.2\\
original & \textbf{96.6} & 95.1 & 92.6 & 94.6& 95.7\\
\bottomrule
\end{tabular}
\end{table*}

\subsection{Confidence metric}
\label{sec:decoding_confidence}
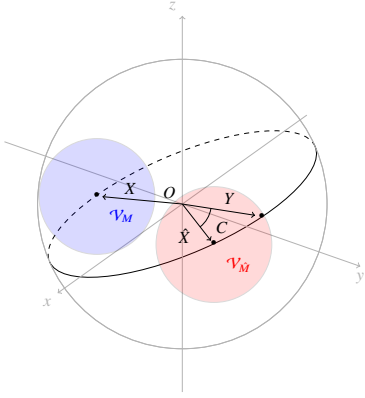
\begin{figure}
    \centering
    \resizebox{0.6\columnwidth}{!}
{
    \centering
  \def\r{3}
  \tdplotsetmaincoords{60}{125}
  
  \begin{tikzpicture}[tdplot_main_coords]
    \draw[tdplot_screen_coords,thin,black!30] (0,0,0) circle (\r);

    \begin{scope}[thin,black!30]
      \tdplotCsDrawGreatCircle%
      [black]{\r}{105}{-23.5}
        \draw[->] (-1.5*\r,0,0) -- (1.5*\r,0,0) node[anchor=north east] {$x$};
        \draw[->] (0,-1.5*\r,0) -- (0,1.5*\r,0) node[anchor=north] {$y$};
        \draw[->] (0,0,-1.5*\r) -- (0,0,1.5*\r) node[anchor=south east] {$z$};
        \draw[tdplot_screen_coords] (0,0,0) circle (\r);
    \end{scope}

    \tdplotsetrotatedcoords{30}{60}{100}
    \draw[fill=red,opacity=0.15] (1,0.5*\r,0) circle (1.20cm) node[below right=0.2cm,opacity = 1.0,font=\small, red] {$\mathcal{V}_{\hat{M}}$};

    \draw[fill=blue,opacity=0.15] (0.3*\r,-1.53,0.1) circle (1.20cm) node[below right=0.2cm,opacity = 1.0,font=\small, blue] {$\mathcal{V}_{M}$};

    

    \draw[fill=black] (0,0) coordinate[label={above left:$O$}] (O);
    
\node  at (0.3*\r,-1.53,0.1) {\textbullet};
\node  at (1,0.5*\r,0) {\textbullet};
\node  at (-0.1710*\r, 0.55*\r, 0.0) {\textbullet};

    \draw[->,] (0,0,0) -- (0.28*\r,-1.43,0.1) 
     coordinate[label={[midway]above left:$X$}] (a);
    \draw[->,] (0,0,0) -- (0.9,0.45*\r,0.0) coordinate[label={[midway]below left:$\hat{X}$}] (b);
    \draw[->,] (0,0,0) -- (-0.16*\r, 0.5*\r, 0.0) coordinate[label={[midway]above right:$Y$}] (c);
    \pic [draw, -, angle eccentricity=1.6, angle radius=0.60cm,"$C$"] {angle=b--O--c};

  \end{tikzpicture}}
    \caption{
Representation of the message space after modulation.
The circles $\mathcal{V}_{M},\mathcal{V}_{\hat{M}}$ illustrate Voronoï cells mapping to different binary messages on the surface of the hyper-sphere. 
$X$ is associated to the message to be hidden $M$.
The extraction retrieves $Y$, which is decoded into $\hat{M}$, while $\hat{X}$ results from the modulation of $\hat{M}$, accounting for the perfect representation of $\hat{M}$.
The similarity between $Y$ and $\hat{X}$ is given by $C$. This provides a confidence score for the decoding as explained in~\secref{sec:decoding_confidence}. The given example illustrates a failure case with a wrong decoded message and a low confidence reflected by a high value of $\rho$.}

    \label{fig:hypersphere}
\end{figure}

After the TCCSK modulation, a message $M$ is encoded into a vector $X$ on the 256-d hypersphere. Given an encoded message $X$, due to transforms during transmission, $Y$ is the noisy extracted vector. At inference, $Y$ is decoded by the TCCSK demodulation into $\hat{M}$. $X$ is unknown, but let $\hat{X}$ be the perfect encoding of $\hat{M}$ and $C$ the cosine similarity computed between $\hat{X}$ and $Y$ (see  \figref{fig:hypersphere}). We define our pratical confidence metric as $\rho$ (\ref{eqn:fisher}). 
We assume that $Y$ close to the perfect representation $\hat{X}$ of the decoded message $\hat{M}$ means that $Y$ is also close to the (unknown) originally embedded vector $X$. Thus, a low $\rho$ would entail a low number of errors at the decoding step, as confirmed by the Pearson correlation coefficient of $-0.89$ computed between $8,000$ $\rho$ values and corresponding MRR. \tabref{tab:wsr_cdf_thresholds} shows the MRR on watermarked images under several scenarios: watermarked images, benign attacks and malign attacks, and three $\rho$-thresholds. Only messages with $\rho < \rho$-threshold should be trusted. An adequate threshold ensures the perfect extraction, meaning $100\%$ MRR, of the watermarked caption.
Note that $\rho$ also refers to the error probability on non-watermarked images: the higher it is, the more a non-watermarked image could be flagged as watermarked. 

A confidence metric could also be used in a multi-bit setting with fixed-length code of 64 bits. Padding a $n$-bit message with $64 - n$ equiprobable random bits drawn from a synchronous source shared by Alice and Bob (e.g. via a secret key $K$) allows to carry confidence information, at the expense of capacity. Indeed, Bob can check the padding bits and discard any message not matching the expected sequence. In this case, assuming the multi-bit system outputs random codes uniformly under $\mathcal{H}_0$, there is still a $2^{-64+n}$ probability of ending up with the expected sequence by chance, giving $\rho \geq 2^{-64} \simeq 5e-20$. Our method benefits from comparable confidence, with $\rho=1.4e-16$ to be protected against all attacks, along with a much greater capacity.

\begin{table}[h]
\centering
\begin{tabular}{c|c|c|c|c}
\textbf{$\rho$ Threshold} & \textbf{Scenario} & \textbf{MRR} & \textbf{CDF} & \textbf{Confidence} \\
\hline
\multirow{3}{*}{1.0} & Watermarked & 96.6 & 100 & \multirow{3}{*}{None} \\
 & Benign Attacks & 89.0 & 100 & \\
 & Malign Attacks & 66.1 & 100 & \\
\hline
\multirow{3}{*}{4.2e-13} & Watermarked & 100 & 82.6 & \multirow{3}{*}{Low} \\
 & Benign Attacks & 96.2 & 60.1 & \\
 & Malign Attacks & 94.1 & 31.1 & \\
\hline

\multirow{3}{*}{2.3e-15} & Watermarked & 100 & 72 & \multirow{3}{*}{Medium} \\
 & Benign Attacks & 100 & 48.3 & \\
 & Malign Attacks & 99.3 & 20.2 & \\

\hline
\multirow{3}{*}{1.4-e16} & Watermarked & 100 & 66.2 & \multirow{3}{*}{High} \\
 & Benign Attacks & 100 & 42.3 & \\
 & Malign Attacks & 100 & 18.1 & \\
\end{tabular}
\caption{Message Recovery Rate (MRR) and Cumulative Distribution Function (CDF) on Emu Edit for different $\rho$ thresholds with SWIFT(40db). Only messages with $\rho < \rho$-threshold are decoded. }
\label{tab:wsr_cdf_thresholds}
\end{table}
\subsection{Qualitative results}
\begin{figure}
    \centering
    \includegraphics[width=\linewidth]{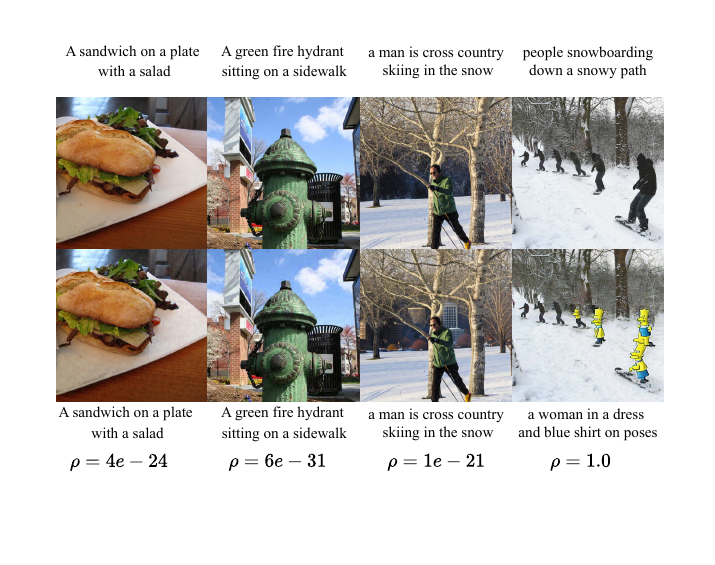}
    \caption{Examples of recovered messages under different transforms. From top to bottom: message, cover image, watermarked image with or without transforms, retrieved message and confidence metric (the lower, the better). From left to right: no transform, jpeg compression with a quality coefficient at 50, background edit, local edit.}
    \label{fig:visual_ex}
\end{figure}

In \figref{fig:visual_ex} we present samples of watermarked images from Emu Edit. In the first column we see an original image and its watermarked version : the watermark is visible in the bottom left on the plate. The second column depicts our resilience against heavy jpeg compression and the two last ones demonstrates SWIFT behaviour against malign edits. The background edit leaves the foreground intact and thus does not destroy the watermark. In the last column, the local edit remove watermarked regions and lead to a low confidence score, suggesting the image is altered and cannot be trusted.

\subsection{Limitations}
We believe this work introduces to a new way of asserting integrity of images. 
By disentangling watermarking and encoding layers, we expose two research directions : better modulation and better representation of the message to transmit. Further research on carrier modulation techniques could potentially enhance performance by reducing inter-carrier interference. 
On the latter, our choice of a text description may be considered simplistic compared to a specifically learnt representation. Moreover, we limit the granularity of captions to reduce the length of the message to encode. This hampers fine-grained comparison but we believe it will be further optimized. 
We leave to future work the task of designing a system taking advantage of our pipeline output: Alice could be considered as the source of the original media while Bob would be a moderation system on a social media platform.

\section{Conclusion}
In this work, we present a novel way to assert the integrity of an image by the relevant use of watermarking as a covert communication channel. Moreover we provide a definition of the image semantics, through its caption, to the recipient of the message. By using an LLM combined with an arithmetic encoder to compress the caption,
the limited capacity of \Hide to convey information is optimized.  

Finally, our local confidence metric improves the applicability of our method as any trusted operator may check if the received image is consistent with the descriptive decoded caption of the original content in a trustworthy manner.

\bibliographystyle{ieeetr}
\bibliography{bib} 

\end{document}